\documentclass[aip, jcp, amsmath,amssymb, reprint]{revtex4-2}

\usepackage{graphicx} 

\usepackage{amsmath}
\usepackage{physics}

\usepackage[utf8]{inputenc}
\usepackage[T1]{fontenc}

\usepackage{etoolbox}

\usepackage{tikz}
\usetikzlibrary{shapes.geometric}
\usetikzlibrary{positioning}
\usepackage{jigsaw}
\usetikzlibrary{math}
\tikzset{
triangle/.style = {fill=blue!20, isosceles triangle, isosceles triangle apex angle=120, draw=blue!20!black,line width=0.5mm, rounded corners=0.1cm},
border rotated/.style = {shape border rotate=270},
fwd_state/.style={ code=
    \node[triangle, border rotated, minimum size=0.8cm] (initial_state) at (0.5,-0.5) {$\rho_0$};
    
        \draw [line width = 0.4mm] 
	(initial_state) -- (0.5,0.25);
	};
   }

\makeatletter
\def\@email#1#2{%
 \endgroup
 \patchcmd{\titleblock@produce}
  {\frontmatter@RRAPformat}
  {\frontmatter@RRAPformat{\produce@RRAP{*#1\href{mailto:#2}{#2}}}\frontmatter@RRAPformat}
  {}{}
}%
\makeatother

\begin{document}

\title{Unifying methods for optimal control in non-Markovian quantum systems via process tensors}

\author{Carlos Ortega-Taberner}
\author{Eoin O'Neill}
\author{Eoin Butler}
\affiliation{School of Physics, Trinity College Dublin, Dublin 2, Ireland}
\affiliation{Trinity Quantum Alliance, Unit 16, Trinity Technology and Enterprise Centre, Pearse Street, Dublin 2, Ireland}
\author{Gerald E. Fux} 
\affiliation{The Abdus Salam International Center for Theoretical Physics (ICTP), Strada Costiera 11, 34151 Trieste, Italy}
\author{P. R. Eastham}
\affiliation{School of Physics, Trinity College Dublin, Dublin 2, Ireland}
\affiliation{Trinity Quantum Alliance, Unit 16, Trinity Technology and Enterprise Centre, Pearse Street, Dublin 2, Ireland}

\date{\today}

\begin{abstract}
  The large dimensionality of environments is the limiting factor in applying optimal control to open quantum systems beyond Markovian approximations. Multiple methods exist to simulate non-Markovian open systems which effectively reduce the environment to a number of active degrees of freedom. Here we show that several of these methods can be expressed in terms of a process tensor in the form of a matrix-product-operator, which serves as a unifying framework to show how they can be used in optimal control, and to compare their performance. The matrix-product-operator form provides a general scheme for computing gradients using back propagation, and allows the efficiency of the different methods to be compared via the bond dimensions of their respective process tensors.

\end{abstract}

\maketitle

\section{Introduction}

Optimal control refers to the problem of determining a set of control fields which can be applied to a dynamical system to induce a particular operation. The application of these techniques to quantum systems~\cite{dong2010,shapiro2011,torrontegui2013,glaser2015,koch2016,koch2019,koch2022} has led to discoveries  ranging from methods to control chemical reactions to techniques for implementing quantum gates and thermodynamic processes. While there have been many such developments, the vast majority of applications refer to the subset of problems that can be treated using a time-local master equation~\cite{breuer2007,jirari2007,chirolli2008,jirari2020}. This excludes consideration of the many important cases of open quantum systems which are coupled, perhaps strongly, to environments with significant spectral structure~\cite{meier1999,ishizaki2005,wenderoth2021}. Such environments are ubiquitous in solid-state and chemical systems, including, for example, the phonon environments experienced by solid-state qubits such as excitons and defect centers; various forms of electromagnetic noise in circuit-QED; and molecular vibrations.  

In this paper, we consider optimal control in open quantum systems, beyond those describable by a time-local master equation. We consider model-based optimal control, where the feedback characterizing the performance of a given control strategy is obtained from a numerical simulation. We restrict ourselves to the typical form of open quantum system model in which a small subsystem of interest, referred to as a `system', is coupled to one or more environments. Most commonly, these are ensembles of harmonic oscillators, although there are other possibilities. The immediate difficulty with such problems is that the system dynamics is affected by the many-particle environment, whose exponentially large state space precludes direct simulation. Nonetheless, various numerical methods are available, including auxiliary-mode methods~\cite{garraway1997,tamascelli2018,basilewitsch2019,pleasance2020}, hierarchical equations-of-motion (HEOM)~\cite{tanimura1989,ishizaki2005,tanimura2006,xu2007,tanimura2020,lambert2023}, and stochastic Liouville equations~\cite{stockburger1999,stockburger2001,stockburger2002, stockburger2004}, all of which effectively reduce the dimensionality of the environment so that it becomes manageable.

An important class of methods derives from the path-integral for an open quantum system~\cite{makri1995,makri1995a,strathearn2018,strathearn2020,pollock2018}, which can be evaluated numerically by taking advantage of the finite memory time of the environment. Recent versions of these techniques take advantage of an efficient matrix product representation of the environment's influence functional~\cite{strathearn2018,strathearn2020,pollock2018}, constructed using truncated singular-value decompositions (SVDs), which allows for simulations with very long memory times. Furthermore, these matrix product representations can be recast in the language of process tensors~\cite{pollock2018} -- objects which provide an operational description of a general quantum stochastic process, or equivalently an open quantum system. 
The process tensor encodes the full effect of the environment, and can be composed with a particular sequence of operations on the system to obtain the system's dynamics. This constitutes a great advantage in the context of optimization~\cite{fux2022,butler2024}. This is because with this approach the process tensor only needs to be calculated once, and can then be applied cheaply in each iteration of an optimization algorithm.

A relevant question when considering optimization of open systems is which of the available methods to simulate non-Markovian open systems one should use. In this paper, we use the process tensor formalism as a way to consider the different methods on the same footing. 
The suitability of each method for implementing optimization can now be assessed by looking at the efficiency in the calculation of the corresponding process tensor, its dimensions, and other advantages or drawbacks.

We discuss the methods for obtaining process tensors for HEOM, stochastic Liouville equations, and auxiliary-mode methods. We also discuss the implementation of optimization using the time-dependent variational principle, and the generalized quantum master equation, for which process tensors cannot be obtained. We compare these methods with those using the Process Tensor in Matrix Product Operator form (PT-MPO) \cite{oqupaper}, where the process tensor is compressed as it is calculated, using singular-value decompositions (SVDs) and truncation.

\section{Background}

\subsection{Optimization and the adjoint method}\label{section:adjoint}

In general, optimization problems involve an objective function, or cost function, that quantifies the quality of a solution, and which one seeks to maximize or minimize over the available control parameters. Methods for doing this follow one of two approaches. Gradient-based approaches rely on computing the gradient of the cost function with respect to the control parameters in order to update them. Some relevant examples in quantum optimal control include Gradient Ascent Pulse Engineering (GRAPE)~\cite{khaneja2005,chen2023,petruhanov2023} and implementations of Krotov algorithms~\cite{reich2012,goerz2019}. On the other hand gradient-free approaches, such as the Chopped Random Basis (CRAB) method~\cite{caneva2011, muller2022}, use a direct-search approach to find the minima of the cost function. While iterations in gradient-free methods are usually more efficient, they converge slowly compared to gradient-based methods, and suffer from lower precision. For these reasons we focus on the gradient-based approach, where the adjoint method~\cite{plessix2006} can be used to efficiently compute the gradient.

We will focus on optimization of terminal costs, as this is an important case in quantum applications. For a system described by variables $\vec{x}(t)$, we consider a terminal cost $Z$ which only depends on the system state at some final time $T$,
\begin{align}
    Z = z(\vec{x}(T)).
\end{align}
The system is subject to the dynamics given by
\begin{align}
     &\dot{\vec{x}}  = \vec{g}(\vec{x}(t),\vec{u}(t)) \nonumber \\
     &\vec{x}(0) = \vec{x}_I,
\end{align}
where $\vec{u}(t)$ is a set of controls, i.e., functions of time that are varied to optimize the process.

\begin{figure}[t]
\centering
\begin{tikzpicture} [
    triangle/.style = {fill=white, isosceles triangle, isosceles triangle apex angle=90, draw=blue!20!black,line width=0.5mm, rounded corners=0.1cm
    },
    border rotated/.style = {shape border rotate=270},
    propagator/.style = {fill=green!20,circle,draw=green!60!black!60,line width=0.4mm},
    mpo/.style = {fill=red!20,rectangle,draw=red!70!black,line width=0.4mm,rounded corners=0.1cm},
    circlepink/.style = {fill=purple!20!white, circle, isosceles triangle apex angle=90, draw=blue!20!black,line width=0.5mm, rounded corners=0.1cm
    },
  ]

    \def \eqx {-0.0}
    \def \eqy {-0.0}

    \node (alabel) at (\eqx-0.7,\eqy+0.8) {$(a)$} ;

    \node (eq1) at (\eqx,\eqy) {$\bra{i}\rho_0 \ket{j} =$} ;
    \node[triangle, border rotated, minimum size=0.02cm] (initial_state_eq) at (\eqx+1.2,\eqy-0.7) {};
    \node (ij) at (\eqx+1.2,\eqy+0.7) {$i,j$};

    \draw[line width = 0.4mm] (initial_state_eq) -- (ij);

    \def \eqxb {4.0}
    \def \eqyb {-0.0}

    \node (eq1b) at (\eqxb,\eqyb) {$\bra{i}U(\rho_0) \ket{j} =$} ;
    \node[triangle, border rotated, minimum size=0.02cm] (initial_state_eqb) at (\eqxb+1.5,\eqyb-0.7) {};
    \node[propagator, minimum size=0.1cm] (propb) at (\eqxb+1.5,\eqyb-0.05) {\tiny \tiny };
    \node (ijb) at (\eqxb+1.5,\eqyb+0.7) {$i,j$};

    \draw[line width = 0.4mm] (initial_state_eqb) -- (propb);
    \draw[line width = 0.4mm] (propb) -- (ijb);


    \def \forx {0.0}
    \def \fory {-6.0}
    
    \def \xa {0.0}
    \def \xb {0.8}
    \def \deltay {0.7}
    
    \node (blabel) at (\forx-0.7,\fory+3.8) {$(b)$} ;
    \node[triangle, border rotated, minimum size=0.02cm, label=left:$\rho_0$] (initial_state) at (\forx+\xb,\fory-0.0) {};

    \foreach \n in {0,1,2,3}{

    \node[propagator, minimum size=0.1cm] (prop\n) at (\forx+\xb,\fory+\deltay*\n+1*\deltay) {\tiny \tiny };

    }

    \draw [line width = 0.4mm] 
    (initial_state) -- (prop0);

    \foreach \n [evaluate=\n as \m using int(\n-1)] in {1,2,3}{
    \draw [line width = 0.4mm] 
    (prop\m) -- (prop\n);
    ;}

    \draw [line width = 0.4mm] 
    (prop3) -- (\forx+\xb,\fory+\deltay*4+\deltay)
    ;
    
    \node[label=left:$\rho_f$] (finalstate) at (\forx+\xb,\fory+\deltay*4+\deltay) {};


    \def \bacx {4.0}
    \def \bacy {-6.0}
    
    \def \xa {0.0}
    \def \xb {0.8}
    \def \deltay {0.7}
    
    \node (blabel) at (\bacx-0.7,\bacy+3.8) {$(c)$} ;
    \node[triangle, fill=blue!20, draw=blue!60!black!60, rotate=180, border rotated, minimum size=0.02cm] (target_state) at (\bacx+\xb,\bacy+\deltay*4+1*\deltay) {};
    \node[label=left:$\lambda_0$] (final_adjoint) at (\bacx+\xb,\bacy-0.0) {};

    \foreach \n in {0,1,2,3}{

    \node[propagator, minimum size=0.1cm] (prop\n) at (\bacx+\xb,\bacy+\deltay*\n+1*\deltay) {\tiny \tiny };

    }

    \draw [line width = 0.4mm] 
    (target_state) -- (prop3);

    \foreach \n [evaluate=\n as \m using int(\n-1)] in {1,2,3}{
    \draw [line width = 0.4mm] 
    (prop\m) -- (prop\n);
    ;}

    \draw [line width = 0.4mm] 
    (prop0) -- (final_adjoint)
    ;
    
    \node[label=left:$\lambda_f$] (finalstate) at (\bacx+\xb-0.1,\bacy+\deltay*4+\deltay) {};
    
\end{tikzpicture}
\caption{ (a) Representation of the density matrix and the time-evolution superoperator in Liouville space. (b) Diagram for the forward propagation of a closed quantum system in Liouville space. (c) Diagram for the propagation of the costate variable from $\lambda_f$ backwards.} \label{fig:closed-diagram}
\end{figure}
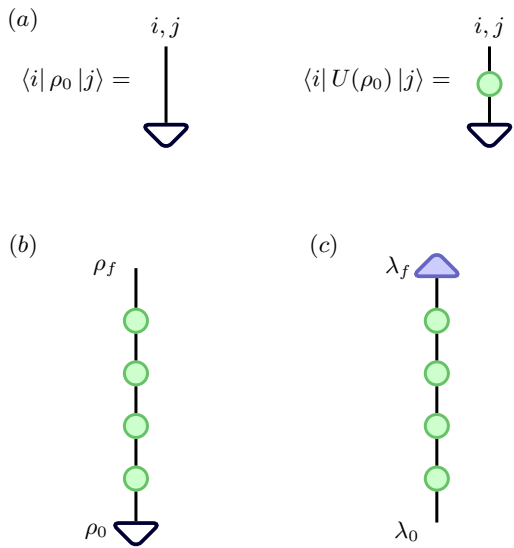

In order to update the control protocol we need to compute the gradient of the cost function with respect to the controls,
\begin{align}
    \frac{\delta Z}{\delta \vec{u}(t)} = \frac{\partial Z}{\partial \vec{x}(T)} \frac{\delta \vec{x}(T)}{\delta\vec{u}(t)}.
\end{align}
The bulk of the complexity in the calculation resides in computing the implicit dependence of the state variables on the controls, $ \delta \vec{x}(T)/\delta\vec{u}(t)$. This calculation is simplified by employing the adjoint method. We will focus on the discretized version of the method, considering the dynamics
\begin{align}\label{eq:adjoint_fwd}
     &\vec{x}_{k}  = \vec{f}(\vec{x}_{k-1},\vec{u}_{k}),\nonumber \\
     &\vec{x}_0 = \vec{x}_I,
\end{align}
where $\vec{u}_k = \vec{u}(t_k)$. The costate variable is now introduced, a Lagrange multiplier whose dynamics are 
\begin{align}
    &  \vec{\lambda}^T_{k-1} = \vec{\lambda}^T_{k}\left.\frac{\partial \vec{f}(\vec{x},\vec{u}_k)}{\partial \vec{x}}\right|_{\vec{x}=\vec{x}_{k-1}}, \label{eq:backprop}\\
    & \vec{\lambda}_{T}^T= -\frac{\partial Z(\vec{x}_T)}{\partial \vec{x}_T}.
\end{align}
This allows the gradient of the cost function to be obtained as
\begin{align}\label{eq:gradient}
    \frac{dZ }{d \vec{u}_k} =&  -\left.\vec{\lambda}_{k+1}^T\frac{\partial \vec{f}(\vec{x}_k,\vec{u})}{\partial \vec{u}}\right|_{\vec{u}=\vec{u}_{k+1}} ,
\end{align}
and used to update the control parameters in a gradient-based optimization.

Let us now look at the computational time cost of implementing the adjoint method. If the computational time to compute $f^\mu(\vec{x}_k,\vec{u}_k)$, for a particular $\mu$, is ${O}(F)$, then propagating the $N$ elements of $\vec{x}$ forward $T$ time steps will take $O(T N F)$. Evaluating the matrix $\frac{\partial \vec{f}}{\partial \vec{x}_k}$ will have, in general, a time cost $O(N^2 F)$, so that the cost of propagating the costate variable backwards using Eq.~\eqref{eq:backprop} over the $T$ time steps is $O(TN^2  F)$. The final step, computing the gradient, has a cost $O(TM N F)$, where $M$ is the number of control parameters per time step. Since typically $M \ll N$, the overall computational cost is dominated by the propagation of the costate variable, $O(T N^2 F)$.

\subsubsection{Closed quantum systems and linearity}

As a simple example we now consider implementing the adjoint method on a closed quantum system. The system is described by its density matrix, $\rho$, which we consider as a vector in Liouville space. We shall make extensive use of tensor network diagrams, in which tensors are represented by nodes, while indices are represented by edges. Joining two edges corresponds to contracting over the corresponding indices. The vectorized density matrix is shown in  Fig.~\ref{fig:closed-diagram}(a). Its evolution is given by
\begin{align}
    \vec{\rho}_{k} = U(\vec{u}_k) \vec{\rho}_{k-1},
\end{align}
where $U$ is the time translation superoperator in Liouville space. Because the evolution is now linear in the density matrix, it can be represented by the tensor network diagram shown in Fig.~\ref{fig:closed-diagram}(b). Furthermore, because of this linearity, the costate variable evolves with the same operator as the system variables, but backwards in time
\begin{align}
    \vec{\lambda}_{k-1}^T =\vec{\lambda}_{k}^T U(\vec{u}_k),
\end{align}
as represented in Fig.~\ref{fig:closed-diagram}(c). For that reason this step is called \emph{back propagation}. Since forward and back propagation steps are now equivalent in cost, calculating the gradient with the adjoint method has an overall cost of $O(TS^4)$, due to the matrix product. Here $S$ is the dimensionality of the system's Hilbert space, so that the density matrix becomes an $S^2$ dimensional vector in Liouville space.

For quantum systems interacting with environments, the large dimensionality of the full system, formed from the small system of interest interacting with a many-body environment, makes it prohibitively expensive to simulate in full, and methods need to be used to obtain an approximate reduced description of the small system. As we will see, in most cases the approximation preserves the linear characteristic of the dynamics, which can be exploited, although this is not always the case.

\subsection{Process Tensors}

The process tensor framework takes an operational approach to the description of open quantum systems \cite{pollock2018}, closely related to quantum combs. For a given initial state, the process tensor is the multilinear map from all possible control sequences, also referred to as interventions, to the final density matrix of the system. The interventions can be unitary transformations, measurements, or, in general, any completely-positive maps. This implies that the process tensor encodes all multi-time correlations and the outcome of any possible experiment on the system.

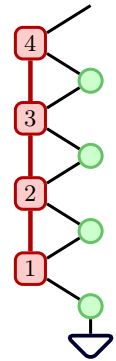
\begin{figure}[t]
\centering
\begin{tikzpicture} [
    triangle/.style = {fill=white, isosceles triangle, isosceles triangle apex angle=90, draw=blue!20!black,line width=0.5mm, rounded corners=0.1cm
    },
    border rotated/.style = {shape border rotate=270},
    propagator/.style = {fill=green!20,circle,draw=green!60!black!60,line width=0.4mm},
    mpo/.style = {fill=red!20,rectangle,draw=red!70!black,line width=0.4mm,rounded corners=0.1cm},
    circlepink/.style = {fill=purple!20!white, circle, isosceles triangle apex angle=90, draw=blue!20!black,line width=0.5mm, rounded corners=0.1cm
    },
  ]
    \def \xa {0.0}
    \def \xb {0.8}
    \def \deltay {1.0}
    
    \node[triangle, border rotated, minimum size=0.02cm] (initial_state) at (\xb,-0.0) {};

    \foreach \n  [evaluate=\n as \m using {int(\n+1)}] in {0,1,2,3}{

    \node[propagator, minimum size=0.1cm] (prop\n) at (\xb,\deltay*\n+0.5*\deltay) {\tiny \tiny };
    \node[mpo,minimum size=0.4cm] (mpo\n) at (\xa,\deltay*\n+\deltay) {\small \m};
    
    }

    \draw [line width = 0.4mm] 
    (initial_state) -- (prop0)
    (prop0) -- (mpo0)
    (mpo1) -- (mpo2) ;

    \foreach \n [evaluate=\n as \m using int(\n-1)] in {1,2,3}{
    \draw [line width = 0.6mm, draw=red!70!black] 
    (mpo\m) -- (mpo\n);
    \draw [line width = 0.4mm]
    (mpo\m) -- (prop\n)
    (prop\n) -- (mpo\n);
    ;}

    \draw [line width = 0.4mm] 
    (mpo3) -- (\xb,\deltay*4+0.5*\deltay)
    ;

\end{tikzpicture}
\caption{Evolution of a system density matrix using a PT-MPO. The connected red squares form the PT-MPO encoding the effect of the environment. The green circles are the system propagators which contain the control parameters. Time increases from the bottom to the top of the diagram.} \label{fig:pt-diagram}
\end{figure}

Let us consider the evolution of a quantum system coupled to an environment by the Hamiltonian
\begin{align}
    H(t) = H_S(\vec{u}(t)) + H_{SE},
\end{align}
where $H_S$ is the system Hamiltonian, with control parameters $\vec{u}$, and $H_{SE}$ captures the environment's Hamiltonian as well as the system-environment interaction. In Liouville space, the evolution of the full density matrix over a time step is given by
\begin{align}
    \ket{\rho_{k}} =& \exp[{i\int_{t_{k-1}}^{t_k} d\tau \, (\mathcal{L}_S(\vec{u}(t)) + \mathcal{L}_{SE})}]\ket{\rho_{k-1}} \nonumber \\
    \approx & e^{i \mathcal{L}_{SE} \Delta t} e^{i \mathcal{L}_S(\vec{u}_k)\Delta t } \ket{\rho_{k-1}} \nonumber \\
    = & F U_k \ket{\rho_{k-1}}.
\end{align}

By evolving multiple times and tracing out the bath degrees of freedom we obtain the density matrix at some final time step $T$, assuming a product state of the system and environment at the initial time, 
\begin{align}
    \ket{\rho_{S,T}} = \left(I \otimes \bra{I}\right) \left(\prod_{k=1}^{T} F U_k\right) \ket{\rho_{S,0}} \otimes\ket{ \rho^{i}_B},
\end{align}
where the projection to $ \left(I \otimes \bra{I}\right) $ corresponds to tracing out the environment in Liouville space. We can introduce a basis for the system so that the components of the vectorized reduced density matrix, at the final time, are
\begin{align}
    \rho^{\mu_{T}}_{S,T} =&\sum_{\{\mu\},\{\nu\}}\left(\bra{I}\prod_{k=1}^{T} F^{\mu_k \nu_k }\ket{\rho^i_B} \right) \left( \prod_{k=1}^{T} U_k^{\nu_k\mu_{k-1}}\right) \rho_{S,0}^{\mu_{0}}  \nonumber \\
    =& \sum_{\{\mu\},\{\nu\}}\mathcal{F}^{\{\mu,\nu\}}\left( \prod_{k=1}^{T} U_k^{\nu_k\mu_{k-1}}\right) \rho_{S,0}^{\mu_{0}}.
\label{eq:pt-state}
\end{align}
Note that $F^{\mu_k\nu_k}$ denotes a matrix element, within the system space, which is an operator in the environment space. $\mathcal{F}$ corresponds to the process tensor, which can be represented in the form of a matrix-product-operator (MPO), 
\begin{align}
\mathcal{F}^{\{\mu,\nu\}} &=  \mathcal{F}^{(\mu_T,\nu_T),...,(\mu_1,\nu_1)} \nonumber \\
&= \sum_{\{\chi\}} O[T]^{\mu_T, \nu_T}_{\chi_T,\chi_{T-1}} 
\cdots O[1]^{\mu_1, \nu_1}_{\chi_1,\chi_{0}},\label{eq:pt-mpoform}
\end{align}
with some bond dimension $\chi_d = \dim(O^{\mu \nu}[k])$. This MPO is  the set of connected squares in Fig.~\ref{fig:pt-diagram}. The indices $\chi_0,\chi_T$ take only a single value, corresponding to the three index tensors at the beginning and end of the MPO. The entire Fig.~\ref{fig:pt-diagram} shows how the PT-MPO can be contracted with a set of system propagators (green circles) and an initial system state to compute the final state, as in Eq.~\eqref{eq:pt-state}.

There are several existing algorithms to obtain the PT-MPO form of the process tensor for various types of environment, as well as software that implements them. The Python package OQuPy~\cite{oqupaper} implements a sequential algorithm introduced by Jørgensen and Pollock \cite{jorgensen2020}, for the case of a Gaussian bosonic environment. The construction is based on representing Eq.\eqref{eq:pt-state} as a two-dimensional tensor network, which is then contracted to form the PT-MPO. As the network is contracted singular-value decompositions are performed, and singular values below some threshold discarded. This limits the exponential growth in bond dimension that would occur if the full Hilbert space of the environment were retained.

Currently, the computation of the process tensor using OQuPy is limited to Gaussian environments. While these represent an important class of problem, it should be noted that the formalism is general, and compressed process tensors for other types of environments can be computed, for example using automatic compression of environments (ACE) \cite{cygorek2021}. Although in the following discussion we will assume $U_k$ to be unitary operations, the method admits any type of interventions on the system, i.e., $U_k$ can describe any completely positive map.

An important advantage offered by the process tensor approach described here is that the calculation of the process tensor needs to be done only once. It can then be used repeatedly to obtain the dynamics for different control sequences. This is particularly relevant for optimization, as one needs to iterate the calculation many times. Note that this is only the case because the control parameters appear only in the system propagators. The method is less suitable if we want, for example, to optimize the coupling strength with the bath. 

\section{Back propagation and optimization using process tensor}\label{section:PT-gradient}

\begin{figure*}[t!]
\centering
\begin{tikzpicture} [
    triangle/.style = {fill=white, isosceles triangle, isosceles triangle apex angle=90, draw=blue!20!black,line width=0.5mm, rounded corners=0.1cm
    },
    border rotated/.style = {shape border rotate=270},
    propagator/.style = {fill=green!20,circle,draw=green!60!black!60,line width=0.4mm},
    mpo/.style = {fill=red!20,rectangle,draw=red!70!black,line width=0.4mm,rounded corners=0.1cm},
    circlepink/.style = {fill=purple!20!white, circle, isosceles triangle apex angle=90, draw=blue!20!black,line width=0.5mm, rounded corners=0.1cm
    },
  ]

    \def \xa {0.0}
    \def \xb {0.8}
    \def \deltay {1.0}

    \node (alabel) at (-2.5,2.5) {$(a)$} ;
    \node (eqx) at (-1.5,1.5) {\Large $\{ \sigma_t\} \rightarrow$};
    
    \node[triangle, border rotated, minimum size=0.02cm] (initial_state) at (\xb,-0.0) {};

    \foreach \n [evaluate=\n as \m using {int(\n+1)}] in {0}{

    \node[propagator, minimum size=0.1cm] (prop\n) at (\xb,\deltay*\n+0.5*\deltay) {\tiny \tiny };
    \node[mpo,minimum size=0.4cm] (mpo\n) at (\xa,\deltay*\n+\deltay) {\small \m};
    
    }

    \draw [line width = 0.4mm] 
    (initial_state) -- (prop0)
    (prop0) -- (mpo0);

    \draw [line width = 0.4mm] 
    (mpo0) -- (\xb,\deltay*1 +0.5*\deltay)
    ;
    \draw [line width = 0.6mm, draw=red!70!black] 
    (mpo0) -- (\xa,\deltay*0.5+\deltay) ;


\begin{scope}[shift={(2.0,0)}]
    \def \xa {0.0}
    \def \xb {0.8}
    \def \deltay {1.0}

    \node (coma) at (-0.7,1.3) {\Large \vphantom{aA},};
    \node (coma) at (\xb+0.7,1.3) {\Large \vphantom{aA}, ...};
    \node[triangle, border rotated, minimum size=0.02cm] (initial_state) at (\xb,-0.0) {};

    \foreach \n [evaluate=\n as \m using {int(\n+1)}] in {0,1}{

    \node[propagator, minimum size=0.1cm] (prop\n) at (\xb,\deltay*\n+0.5*\deltay) {\tiny \tiny };
    \node[mpo,minimum size=0.4cm] (mpo\n) at (\xa,\deltay*\n+\deltay) {\small \m};
    
    }

    \draw [line width = 0.4mm] 
    (initial_state) -- (prop0)
    (prop0) -- (mpo0);

    \foreach \n [evaluate=\n as \m using int(\n-1)] in {1}{
    \draw [line width = 0.6mm, draw=red!70!black] 
    (mpo\m) -- (mpo\n);
    \draw [line width = 0.4mm]
    (mpo\m) -- (prop\n)
    (prop\n) -- (mpo\n);
    ;}

    \draw [line width = 0.4mm] 
    (mpo1) -- (\xb,\deltay*2 +0.5*\deltay)
    ;
    \draw [line width = 0.6mm, draw=red!70!black] 
    (mpo1) -- (\xa,\deltay*1.5+\deltay) ;

\end{scope}
\begin{scope}[shift={(0.0,-3.5)}]
    \def \xa {0.0}
    \def \xb {0.8}
    \def \deltay {1.0}

    \node (blabel) at (-2.5,2.5) {$(b)$} ;
    \node (eqx) at (-1.5,1.5) {\Large $\{ \lambda_t\} \rightarrow$};
    
    \node[triangle, fill=blue!20, draw=blue!60!black!60, rotate=180, border rotated, minimum size=0.02cm] (final_state) at (\xb,\deltay*1+0.5*\deltay+0.3) {};
    
    \node[propagator, minimum size=0.1cm] (prop4) at (\xb,\deltay*0+0.5*\deltay) {\tiny \tiny };
    \node[mpo,minimum size=0.4cm] (mpo4) at (\xa,\deltay*0+\deltay) {\small 4};
    \node[mpo,minimum size=0.4cm, fill=white,draw=white] (mpo3) at (\xa,\deltay*-1+\deltay) {\small };
    
    \draw [line width = 0.4mm] 
    (final_state) -- (\xb,\deltay*1+0.5*\deltay)
    (mpo4) -- (\xb,\deltay*1+0.5*\deltay)
    (prop4) -- (mpo3)
    (prop4) -- (mpo4);

    \draw [line width = 0.6mm, draw=red!70!black] 
    (mpo4) -- (\xa,0.25*\deltay) ;


\begin{scope}[shift={(2.0,0)}]
    \def \xa {0.0}
    \def \xb {0.8}
    \def \deltay {1.0}

    \node (coma) at (-0.7,1.3) {\Large \vphantom{aA},};
    \node (coma) at (\xb+0.7,1.3) {\Large \vphantom{aA}, ...};

    \node[triangle, fill=blue!20, draw=blue!60!black!60, rotate=180, border rotated, minimum size=0.02cm] (final_state) at (\xb,\deltay*1+0.5*\deltay+0.3) {};
    
    \node[propagator, minimum size=0.1cm] (prop4) at (\xb,\deltay*0+0.5*\deltay) {\tiny \tiny };
     \node[propagator, minimum size=0.1cm] (prop3) at (\xb,\deltay*-1+0.5*\deltay) {\tiny \tiny };
    \node[mpo,minimum size=0.4cm] (mpo4) at (\xa,\deltay*0+\deltay) {\small 4};
    \node[mpo,minimum size=0.4cm] (mpo3) at (\xa,\deltay*-1+\deltay) {\small 3};
    \node[mpo,minimum size=0.4cm, fill=white,draw=white] (mpo2) at (\xa,\deltay*-2+\deltay) {\small };
    
    \draw [line width = 0.4mm] 
    (final_state) -- (\xb,\deltay*1+0.5*\deltay)
    (mpo4) -- (\xb,\deltay*1+0.5*\deltay)
    (prop4) -- (mpo3)
    (mpo3) -- (prop3)
    (prop4) -- (mpo4)
    (prop3) -- (mpo2);

    \draw [line width = 0.6mm, draw=red!70!black] 
    (mpo4) -- (mpo3) 
    (mpo3) -- (mpo2);

\end{scope}
\end{scope}
\begin{scope}[shift={(6.5,-3.0)}]
 \def \xa {0.0}
    \def \xb {0.8}
    \def \deltay {1.0}

    \node (eq) at (-1.0,2.5) {\Large $\frac{dZ }{d{
    U}_3} =$};
    
    \node[triangle, border rotated, minimum size=0.02cm] (initial_state) at (\xb,-0.0) {};
    \node (clabel) at (-2.0,5.0) {$(c)$} ;
    \node[triangle, fill=blue!20, draw=blue!60!black!60, rotate=180, border rotated, minimum size=0.02cm] (final_state) at (\xb,\deltay*4+0.5*\deltay+0.3) {};
    
    \foreach \n [evaluate=\n as \m using {int(\n+1)}] in {0,1,3}{

    \node[propagator, minimum size=0.1cm] (prop\n) at (\xb,\deltay*\n+0.5*\deltay) {\tiny \tiny };
    \node[mpo,minimum size=0.4cm] (mpo\n) at (\xa,\deltay*\n+\deltay) {\small \m};
    
    }
    \node[mpo,minimum size=0.4cm] (mpo2) at (\xa,\deltay*2+\deltay) {\small 3};
    \node[propagator, minimum size=0.1cm, fill=white, draw=white] (prop2) at (\xb,\deltay*2+0.5*\deltay) {\tiny \tiny };
    \draw [line width = 0.4mm] 
    (initial_state) -- (prop0)
    (prop0) -- (mpo0)
    (mpo1) -- (mpo2) ;

    \foreach \n [evaluate=\n as \m using int(\n-1)] in {1,2,3}{
    \draw [line width = 0.6mm, draw=red!70!black] 
    (mpo\m) -- (mpo\n);
    \draw [line width = 0.4mm]
    (mpo\m) -- (prop\n)
    (prop\n) -- (mpo\n);
    ;}

    \draw [line width = 0.4mm] 
    (mpo3) -- (\xb,\deltay*4+0.5*\deltay)
    (final_state) -- (\xb,\deltay*4+0.5*\deltay)
    ;

\end{scope}
\end{tikzpicture}
\caption{In implementing the adjoint method using the PT-MPO formalism (a) we identify the system variables as the extended states $\sigma$ and (b) the costate variable $\lambda$ is obtained by back propagation from a final state. (c) All the different diagrams are generated and then stitched with a corresponding additional tensor to obtain the derivative of the cost function with respect to each propagator. This is then used to calculate the derivative of the cost function with respect to the control parameters.} \label{fig:opt-diagram}
\end{figure*}
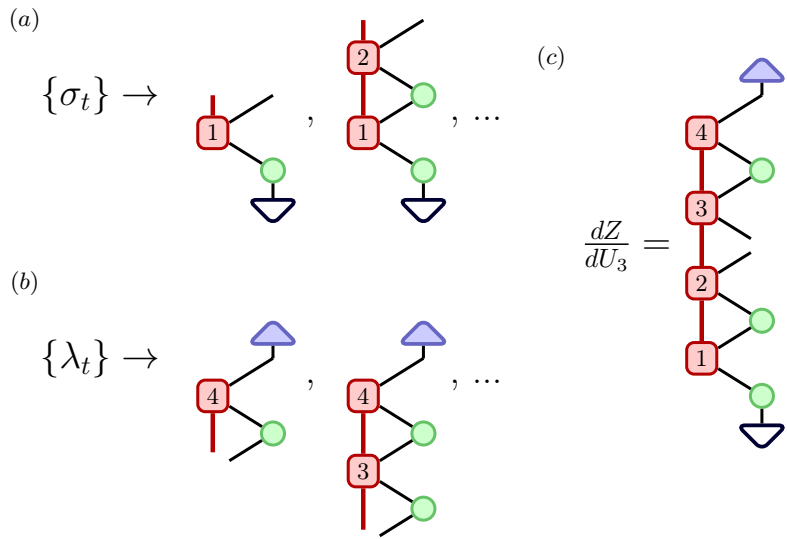

\subsection{Summary of approach}

Looking at the tensor diagram in Fig. \ref{fig:pt-diagram}, we can reinterpret the dynamics as the evolution of an \emph{extended state}
\begin{align}
\sigma^{\mu_k \chi_k}_{k} &=
\sum_{\nu_k,\mu_{k-1}} \sum_{\chi_{k-1}} O[k]_{\chi_k, \chi_{k-1}}^{\mu_k, \nu_k} U^{\nu_k, \mu_{k-1}}_k \sigma^{\mu_{k-1} \chi_{k-1}}_{k-1} \nonumber\\
&= f_k^{\mu_k \chi_k}(\vec{u}_{k},{\sigma}_{k-1}),
\end{align}
with  $\sigma^{\mu_0,\chi_0}_0=\rho_{S,0}^{\mu_0}$, where $\mu_k,\chi_k$ represent the system index and the index of the bond dimension, respectively, with dimensions $(S^2,\chi_d)$. Note that the bond dimension in the extended state is one for $k=0,T$ and we can omit the leg. The extended states are the two-legged objects depicted in Fig.~\ref{fig:opt-diagram}(a). The physical state at each time step can be obtained from the extended state~\cite{cygorek2022}. Taking the extended state to represent the system variable, $\vec{x}$, we continue the adjoint method of Sec.~\ref{section:adjoint} by defining a costate variable $\lambda^{\mu_k \chi_{k}}_{k}$ such that 
\begin{align}
&\lambda^{\mu_{k-1}\chi_{k-1}}_{k-1} = \sum_{\mu_{k},\nu_{k-1}} \sum_{\chi_{k}} \lambda^{\mu_{k}\chi_{k}}_{k}  O[k]^{\mu_{k}\nu_{k-1}}_{\chi_k \chi_{k-1}} U_{k}^{\nu_{k-1}\mu_{k-1}}    \nonumber \\
&\lambda^{\mu_T \chi_T}_{T} =  \left[ \frac{dZ}{d\vec{\rho}_S(T)} \right]^{\mu_T} ,
\end{align}
where $d Z/d \vec{\rho}$ is the gradient of the cost function at the final step. The costate variable, also referred to as the back propagated state, is depicted diagramatically in Fig.~\ref{fig:opt-diagram}(b). Following Eq.~\eqref{eq:gradient},
the gradient of the cost function with respect to the propagator at step $q\leq T$ is then
\begin{align}
\left[\frac{\partial Z}{\partial U_q}\right]^{\mu_{q-1} \nu_q} 
&= \sum_{\mu_q,\chi_q,\chi_{q-1}}
\lambda_{q}^{\mu_q\chi_{q}} O[q]^{\mu_q \nu_q}_{\chi_q \chi_{q-1}}\sigma^{\mu_{q-1} \chi_{q-1}}_{q-1}, 
\end{align}
which corresponds to Fig.~\ref{fig:opt-diagram}(c) for $q=3$ and $T=4$. The derivative of the cost function with respect to the control parameters is then obtained using the chain rule 
\begin{align}
\frac{\partial Z}{\partial u_q^\gamma} = \sum_{\alpha, \beta} \left[\frac{\partial Z}{\partial U_q}\right]^{\alpha\beta}  \frac{\partial U_q^{\alpha\beta}}{\partial u_q^\gamma} ,
\end{align}
which is equivalent to inserting the derivative of the propagator with respect to the control parameter between the open legs in Fig.~\ref{fig:opt-diagram}(c). In practice, $\partial U_q/\partial \vec{u}_q$ is calculated using a finite-difference method.

This method is implemented in the OQuPy package, by generating the extended states and costate variables iteratively, and then contracting the appropriate tensors to generate the gradient. Details of the implementation and usage of the method are given in Ref.~\onlinecite{oqupaper}.

\subsection{Numerical complexity in terms of bond dimension}
\label{sec:ptcomplexity}

We now look at the computational cost of implementing the adjoint method on the process tensor. We will assume, based in our experience, that creating the process tensor is less costly than implementing optimization, neglecting its contribution.
The extended states have dimension $\textrm{dim}[\vec{\sigma_k}] = S^2\chi_d$, where $S$ is the Hilbert-space dimension of the system, and $\chi_d$ is the bond dimension of the PT-MPO. In constructing the PT-MPO, we usually find that best performance is achieved when fixing the relative SVD truncation threshold, which, however, leads to a a bond dimension which varies along the MPO. For the sake of simplicity, here we assume a constant bond dimension, which can be taken as the largest one. Since $S \ll \chi_d$, the cost of contracting the tensors in the MPO with the propagators can be neglected, and the forward propagation step has a time cost of $O(T S^4 \chi_d^2)$. Due to the linearity of the tensor network, back propagation amounts to solving the same equations backwards, and therefore has the same cost as forward propagation, so that the overall time cost remains $O(T S^4 \chi_d^2)$.

\section{Comparison with other approaches}

In the literature of open systems one can find multiple other approaches that have been used to compute the dynamics of a system coupled to a structured environment, accounting for non-Markovian effects. In this section we will summarize some of the most common ones.

Most of these methods are based on identifying a subspace of the environment which is small enough to be simulated yet encompasses the physics of interest. This is in line with the techniques described above, but differs in that the identification is generally made using physical insight rather than a numerical technique. Nonetheless, due to this connection, most of these methods can be related to the process tensor formalism. This reduces the problem of implementing optimal control using these different methods to computing the respective PT from each method, and implementing optimal control as described in Sec.~\ref{section:PT-gradient}. This allows us to compare the efficiency in implementing optimal control by simply comparing the bond dimensions in the respective process tensors. 

In the few cases where a process tensor cannot be directly obtained we analyze the cost of implementing the adjoint method using that particular approach. We also discuss other advantages and drawbacks of the methods when relevant.

\subsection{Time-dependent variational principle}

The simplest way of accounting for the memory effect of a structured environment is to treat system and environment as a closed system, using approximations to efficiently handle the large number of degrees of freedom. One method that has been used to simulate large dimensional systems, including open quantum systems, is the time-dependent variational principle (TDVP) \cite{kramer1981,broeckhove1988,hackl2020}. 

As the starting point, consider a variational ansatz for the system plus environment state $\ket{\Psi(\vec{x})}$,  where $\vec{x}$ is, without loss of generality, a vector of real variational parameters. TDVP approximates the dynamics of the Schr\"odinger equation within the variational manifold, resulting in a set of equations of motion (EOM) for the variational parameters
\begin{align}
     \dot{x}_\mu  =\Omega_{\nu \mu}(\vec{x}) \partial_\nu  \bra{\psi(\vec{x})} H(\vec{u}) \ket{\psi(\vec{x})},
\end{align}
where
\begin{align}
    \Omega^{-1}_{\mu \nu}(\vec{x}) = i\bra{\partial_\mu \Psi(\vec{x})}\ket{\partial_\nu \Psi(\vec{x})}
\end{align}
is the geometric curvature.  Because the EOMs are non-linear with respect to the variational parameters, one cannot disentangle the control parameters from the effect of the bath. Therefore this method cannot be expressed in the process tensor formalism. Nonetheless, after discretizing the EOMs, the adjoint method can be directly applied.

The efficiency of the method depends heavily on how good the ansatz is at approximating the real dynamics while keeping the number of variational parameters small. There are mainly two approaches when applying TDVP, constructing general ansatzes systematically, or using physically informed ansatzes.

\begin{figure}[t]
\centering

\begin{tikzpicture} [
    triangle/.style = {fill=white, isosceles triangle, isosceles triangle apex angle=90, draw=blue!20!black,line width=0.5mm, rounded corners=0.1cm
    },
    border rotated/.style = {shape border rotate=270},
    propagator/.style = {fill=blue!20,rectangle,draw=blue!60!black!60,line width=0.4mm,scale=2},
    mpo/.style = {fill=red!20,rectangle,draw=red!70!black,line width=0.4mm,rounded corners=0.1cm},
    circlepink/.style = {fill=purple!20!white, circle, isosceles triangle apex angle=90, draw=blue!20!black,line width=0.5mm, rounded corners=0.1cm
    },
  ]
    \def \forx {0.0}
    \def \fory {0.0}
    
    \def \xa {0.0}
    \def \xb {0.8}
    \def \deltay {1.0}

    \node[triangle, border rotated, minimum size=0.02cm, label=left:$\vec{x}_0$] (initial_state) at (\forx+\xb,\fory-0.0) {};

    \foreach \n  [evaluate=\n as \m using int(\n+1)] in {0,1,2}{

    \node[propagator, minimum size=0.1cm,label=left:{$f(\vec{x}_\n,\vec{u}_{\m})$}] (prop\n) at (\forx+\xb,\fory+\deltay*\n+1*\deltay) {\tiny \tiny };

    }

    \draw [line width = 0.4mm] 
    (initial_state) -- (prop0);

    \foreach \n [evaluate=\n as \m using int(\n-1)] in {1,2}{
    \draw [line width = 0.4mm] 
    (prop\m) -- (prop\n);
    ;}

    \draw [line width = 0.4mm] 
    (prop2) -- (\forx+\xb,\fory+\deltay*3+\deltay)
    ;
    
    \node[label=left:$\vec{x}_f$] (finalstate) at (\forx+\xb,\fory+\deltay*3+\deltay) {};

\end{tikzpicture}
\caption{Diagram depicting the propagation of variational parameters in TDVP. Note that the variables propagate in a non-linear way.} \label{fig:tdvp-diagram}
\end{figure}
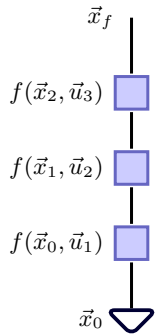
\subsubsection{Systematic ansatzes: MPS and MCTDH}

In the first approach one constructs the ansatz by using a discrete basis which is relevant for the problem, but rather general. One example of this is the application of TDVP to matrix product states (MPS) \cite{haegeman2011,haegeman2016}, widely used on strongly-correlated systems. An MPS is written as
\begin{align}
    \ket{\Psi} = \sum_{\vec{s}} \Tr[A_1^{s_1} A_2^{s_2} ... A_N^{s_N}]\ket{s_1 s_2 ... s_N},
\end{align}
where $A_j^{s_j}$ are complex matrices forming a tensor train, and $\ket{s_1 s_2 ... s_N}$ is a basis for the closed system. An efficient representation is obtained by finding the matrices $A_j^{s_j}$ that best approximate the physical state with the lowest dimension possible. The matrices are then used  as variational parameters to apply TDVP.

Another approach to constructing general ansatzes was introduced by the multi-configuration time-dependent Hartree (MCTDH) method \cite{meyer2009,mundt2009,wang2015}, which has been widely used in physical chemistry. In MCTDH one constructs an ansatz using a time-dependent basis,
\begin{align}
    \ket{\psi(t)} = \sum_{j_1,..,j_f} A_{j_1...j_f}(t) \ket{\phi_{j_1}^1 (t)} \otimes ... \otimes \ket{\phi_{j_f}^f (t)},
\end{align}
which is in turn defined in terms of a time-independent single-particle basis
\begin{align}
    \ket{\phi_{j_k}^k (t)} = \sum_{i_k}c_{i_k j_k}^k(t) \ket{\chi_{i_k}^k}.
\end{align}
TDVP can then be applied using both $A_{j_1...j_f}$ and $c_{i_k j_k}^k$ as variational parameters. 

Both MPS and MCTDH have been used in the study of open systems, particularly solving the spin-boson model \cite{yao2013,schroder2016,finsterholzl2020,wenderoth2021,kamar2023}. Since these methods solve both system and environment, they provide direct access to the dynamics of the structured bath, making them are a very interesting avenue to further understand non-Markovianity. However, for many applications in optimal control only the effect of the bath on the system is relevant, and not its internal dynamics, making these methods excessive compared to other methods developed specifically for this purpose, such as PT-MPO.

\subsubsection{Physically informed ansatzs}

By enlarging the variational manifold one can improve the accuracy of the simulation, but the problem quickly becomes unmanageable. Rather than using general ansatzes as described above, another approach is to use our knowledge of the system to write ansatzes that reduce the complexity of the problem to a few relevant degrees of freedom. To illustrate this approach, we consider the spin-boson model, defined by the Hamiltonian
\begin{align}\label{eq:sb-model}
    H = \frac{\omega_q}{2} \sigma_x + \sum_k \omega_k b_k^\dagger b_k + \frac{\sigma_z}{2} \sum_k g_k (b_k^\dagger + b_k).
\end{align}
The ground state is known to be well approximated by a polaron ansatz \cite{chin2011,bera2014,xu2016,hsieh2019}
\begin{align}
    \ket{\Psi(\vec{x}\,)} = \frac{1}{\sqrt{2}}\left( \ket{\uparrow,\vec{x}} - \ket{\downarrow,-\vec{x}} \right),
\end{align}
where $\ket{\vec{x}}$ is a coherent state of the oscillators displaced by $\vec{x}$. By promoting the displacement to a time-dependent variable and using TDVP, we can approximate the relaxation from an initial separable state $\ket{\Psi(0)} = \ket{-}\otimes\ket{0}$. This results in the EOMs 
\begin{align}
    \frac{d x_k}{dt} =& i x_k(\omega_q e^{-2\sum_p \abs{x_p}^2}+\omega_k) + i\frac{1}{2}g_k ,
\label{eq:polaroneoms}\end{align}
and the magnetization
\begin{align}
    \expval{\sigma_x} =& - e^{-2\sum_k \abs{x_k}^2 }.
\end{align}

\begin{figure}[t]
\centering
\includegraphics[width=\columnwidth]{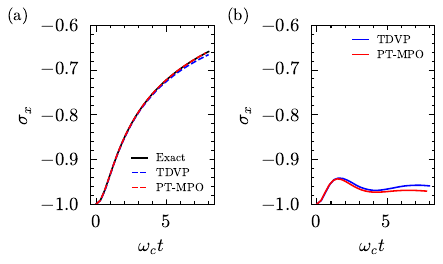}
\caption{(a) Magnetization for the independent boson model with an Ohmic bath and an exponential cutoff, with $\alpha = 0.1$. We compare the result using the TDVP method with the polaron ansatz to the PT-MPO and exact results. (b) Magnetization for the spin-boson model with $\omega_q = \omega_C$. In both cases the degrees of freedom of the environment (bond dimension for PT-MPO and number of oscillators for TDVP) are kept to $ \simeq 30$.} \label{fig:tdvp-res}
\end{figure}

Despite the simplicity of the ansatz, and with just $N=30$ variational parameters in the calculation, TDVP reproduces very well the dynamics in the $\omega_q=0$ case, where an exact solution is known.  This is is shown in Fig.~\ref{fig:tdvp-res}(a). For finite $\omega_q$ we compare, in Fig \ref{fig:tdvp-res}(b), the results from TDVP with a converged result using the PT-MPO method implemeted in OQuPy. While the deviation of the TDVP result is larger at longer times, it still reproduces the qualitative behavior of the dynamics. Note that increasing the number of variational parameters only improves the result slightly, as the error comes mainly from the ansatz being too restrictive. 

\begin{figure}[t]
\centering
\includegraphics[width=\columnwidth]{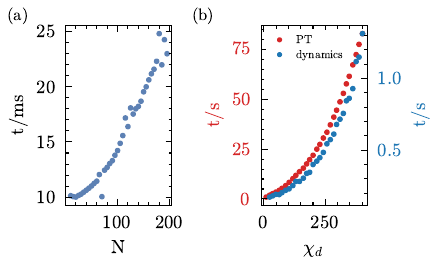}
\caption{(a) Scaling cost of calculating the gradient using the adjoint method for a TDVP calculation. This shows a quadratic behavior due to the structure of the differential equations. (b) Scaling cost of creating the process tensor and calculating the dynamics using OQuPy. Creating the process tensor is found to scale linearly at small bond dimensions, with a cubic scaling at larger bond dimensions, where the calculation is dominated by the SVD step. Computing the dynamics is dominated by the tensor contraction, which has a quadratic scaling. Back propagation has the same quadratic scaling in this case.} \label{fig:tdvp-scaling}
\end{figure}

For $\omega_q \neq 0$ the EOMs need to be solved numerically. In general, TDVP will result in a set of equations in which the dynamics of each variational parameter depends on every other variational parameter. Thus, the cost of evaluating the right-hand side of one of the dynamical equations, defined to be $O(F)$ in Sec.~\ref{section:adjoint}, will be at least $O(N)$ in terms of the number of variational parameters. Thus the time complexities for forward and back propagation become $O(TN^2)$ and $O(TN^3)$, respectively.  However, the EOMs from the polaron ansatz, Eq.~\eqref{eq:polaroneoms}, have a special form
\begin{align}
    \frac{d x_k}{dt} = x_k A + B
\end{align}
where $B$ is independent of the state variables, and $A$ depends on a single function of them, that therefore needs to be computed only once per time step. This reduces $O(F)$ to being $O(1)$ in terms of the number of variational parameters, as for the linear case, and the forward and backward propagation costs to $O(TN)$ and $O(TN^2)$ respectively. This scaling of the backward propagation cost is confirmed by our numerical results in Fig. \ref{fig:tdvp-scaling}(a). The same $O(N^2)$ scaling is found in the calculation of the dynamics using the PT-MPO method in OQuPy, as shown in Fig.~\ref{fig:tdvp-scaling}(b), in line with our predictions in Sec.~\ref{sec:ptcomplexity}. 

It can be seen from Fig.~\ref{fig:tdvp-scaling} that, with this simple ansatz, our implementation of propagation using TDVP is faster than that using the PT-MPO, for $N$ similar to the bond dimension $\chi_d$. However, in order to obtain more accurate results, one needs a more complex ansatz, which will result in a larger cost. In general, such an ansatz would bring the scaling back to $O(N^3)$.

The simplicity of the physically informed ansatzes can therefore be exploited to reduce computational costs. While the ansatz described here as an example is too restrictive, extensions of the polaron ansatz have been shown to approximate the spin-boson model quite well. That is the case for the family of Davydov ansatzes\cite{yao2013,wang2016,chen2018}. Furthermore, we only discussed ansatzes for pure states, which do not include effects such as temperature. Implementing TDVP directly on density matrices is problematic as it results in dynamics where the trace of the density matrix, as well as the energy, are not conserved \cite{joubert-doriol2015}. While there are extensions of the method that fix these issues \cite{joubert-doriol2015,paeckel2019}, they are less straightforward to implement.

For the example described above, TDVP is a good method for implementing optimal control due to its simplicity. However, the reliance on knowing a good ansatz and the uncontrollability of the approximation make the method less likely to be useful in the general case. 

\subsection{Generalized quantum master equation}

In the literature one can find generalizations of the master equation that include non-Markovian effects. One of the most relevant includes a memory kernel\cite{chruscinski2022,chruscinski2010,breuer2008,vacchini2016,cerrillo2014,gherardini2022,jorgensen2020,pollock2018,ivander2024} in the master equation, as 
\begin{align}
    \partial_t \rho (t) = \mathcal{L}_S(t) \rho(t) + \int_0^t d\tau \, K(t,\tau) \rho(\tau),\label{eq:nakazwan}
\end{align}
where the memory kernel $K(t,\tau)$ encodes the memory effects of the environment. 

The dynamics of any reduced density matrix can be expressed as above using the Nakajima-Zwanzig formalism \cite{reineker1980,grabert1982,shi2003,zhang2006,xu2018}, which may lead one to suspect it provides a general foundation for optimal control beyond the Markovian limit. This is not, however, the case, because the evolution of the density operator is not sufficient to determine multi-time correlations~\cite{oqupaper,milz2021}, which are the key object in optimal control. To explain this last point, consider the problem of choosing a time-dependent Hamiltonian, contained in the propagators in Fig.~\ref{fig:pt-diagram}, to reach some state $\ket{\sigma}$. The probability of success is the multi-time correlation function, $\langle \ket{\sigma}\bra{\sigma} U_4 U_3 U_2 U_1\rangle$, where $\langle\rangle$ denotes the average over the environment and system. Thus, quantum optimal control demands a description of the environment in terms of multi-time correlation functions. This information is captured by the process tensor, but not necessarily by the Nakajima-Zwanzig memory kernel. This implies that one can obtain the memory kernel from the process tensor\cite{pollock2018,ivander2024}, but not the reverse.

Despite this, one could implement optimal control using Eq. \eqref{eq:nakazwan}. The difficulty is that the memory kernel depends on the system Hamiltonian, and so would need to be computed, for example using perturbation theory, for every different control. This will make the method inherently less efficient than those which use the process tensor. An exception occurs if the memory kernel is independent of the system Hamiltonian, as it is, for example, for the independent boson model\cite{chin2012}. In the literature one can also find phenomenological models for the memory kernel where it is taken to be system independent \cite{chruscinski2010,breuer2008,vacchini2016}. While this would greatly simplify the calculation, the assumption is only correct when the noise experienced by the system comes from an external source, and not for a system coupled to an environment\cite{addis2016}. 

In light of these exceptions, we consider 
the computational difficulty of implementing the adjoint method and optimal control using Eq. \eqref{eq:nakazwan}, assuming a fixed memory kernel. It can be rewritten as \begin{align}
    \rho_{k+1} = \sum_{j=0}^k T_{k j} \rho_j,\label{eq:transfertensor}
\end{align}
where $T_{kj}$ is known as the transfer tensor\cite{pollock2018,cerrillo2014,jorgensen2020,gherardini2022}. We can implement a memory cutoff $T_C$, such that $T_{j+T_C,j} = 0$. Then the cost of using Eq. (\ref{eq:transfertensor}) to advance $T$ time steps will be $O(T T_C S^4)$ in the long time limit $T \gg T_C$. Since the propagation is linear in the density matrix, the adjoint method will have this same time complexity.

\subsection{Hierarchical equations of motion}

The method of Hierarchical Equations Of Motion (HEOM) is an exact approach that has been very successful in the study of structured environments \cite{tanimura1989,ishizaki2005,tanimura2006,xu2007,tanimura2020,lambert2023}, and has  been applied in quantum optimal control \cite{mangaud2018}. It relies on approximating the bath correlation function by a finite series of exponentials,
\begin{align}
    &C(t-\tau) = \sum_{j=1}^M \alpha_j e^{i \gamma_j (t-\tau)} , \nonumber\\
    &C^*(t-\tau) = \sum_{j=1}^M \tilde{\alpha}_j e^{i \gamma_j (t-\tau)},
\end{align}
where the parameters $\alpha_k, \tilde{\alpha}_k$ and $\gamma_k$ are to be obtained by fitting the bath spectral function. Expanding the Liouville equation and identifying recurring terms, one can obtain a set of coupled equations of motion for auxiliary density matrices, $\rho^{\vec{n}}$. In Liouville space, the coupled equations of motion are
\begin{align}
    \partial_t \ket{\rho^{\vec{n}}} =& \mathcal{L}_S \ket{\rho^{\vec{n}}} + i \sum_{k=1}^M n_k \gamma_k \ket{\rho^{\vec{n}}} \nonumber\\
    &- i \sum_{k=1}^M \left( S \otimes I - I \otimes S \right) \ket{\rho^{\vec{n} + \vec{u}_k}} \nonumber\\
   & - i \sum_{k=1}^M n_k \left( \alpha_k S \otimes I - I \otimes \tilde{\alpha}_k S \right) \ket{\rho^{\vec{n} - \vec{u}_k}},
\end{align}

\begin{figure}[t]
\centering
\begin{tikzpicture} [
    triangle/.style = {fill=white, isosceles triangle, isosceles triangle apex angle=90, draw=blue!20!black,line width=0.5mm, rounded corners=0.1cm
    },
    border rotated/.style = {shape border rotate=270},
    propagator/.style = {fill=green!20,circle,draw=green!60!black!60,line width=0.4mm},
    mpo/.style = {fill=red!20,rectangle,draw=red!70!black,line width=0.4mm,rounded corners=0.1cm},
    circlepink/.style = {fill=purple!20!white, circle, isosceles triangle apex angle=90, draw=blue!20!black,line width=0.5mm, rounded corners=0.1cm
    },
    half circle/.style={fill=white,semicircle,shape border rotate=180,anchor=chord center,  draw=blue!20!black,line width=0.5mm, rounded corners=0.02cm
      }
  ]
    \def \xa {0.0}
    \def \xb {0.8}
    \def \deltay {1.0}
    
    \node[triangle, border rotated, minimum size=0.02cm, label=right:{$\rho_0$}] (initial_state) at (\xb,-0.0) {};
    \node[half circle, minimum size=0.02cm, label=left:{\scriptsize $\bra{\vec{0}}$}] (initial_cap) at (\xa,0.5) {};
    \node[half circle, rotate=180,minimum size=0.02cm, label=right:{\scriptsize $\ket{\vec{0}}$}] (final_cap) at (\xa,\deltay-0.5+\deltay*3+\deltay) {};

    \foreach \n in {0,1,2,3}{

    \node[propagator, minimum size=0.1cm] (prop\n) at (\xb,\deltay*\n+0.5*\deltay) {\tiny \tiny };
    \node[mpo,minimum size=0.4cm,label = left:{$O$}] (mpo\n) at (\xa,\deltay*\n+\deltay) {\small };
    
    }

    \draw [line width = 0.4mm] 
    (initial_state) -- (prop0)
    (initial_cap) -- (mpo0)
    (prop0) -- (mpo0)
    (mpo1) -- (mpo2) ;

    \foreach \n [evaluate=\n as \m using int(\n-1)] in {1,2,3}{
    \draw [line width = 0.6mm, draw=red!70!black] 
    (mpo\m) -- (mpo\n);
    \draw [line width = 0.4mm]
    (mpo\m) -- (prop\n)
    (prop\n) -- (mpo\n);
    ;}

    \draw [line width = 0.4mm] 
    (mpo3) -- (\xb,\deltay*4+0.5*\deltay)
    (mpo3) -- (final_cap)
    ;

\end{tikzpicture}
\caption{Diagram of the PT-MPO implementation from HEOM. The MPOs at different times are now all equal. The final process tensor is obtained by projecting the initial and final MPO on the empty state $\ket{\vec{0}}$. } \label{fig:heom-diagram}
\end{figure}
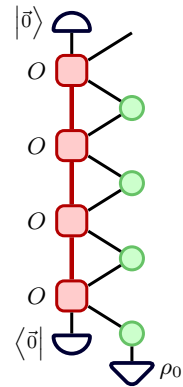
together with the initial conditions
\begin{align}
    &\ket{\rho^{\vec{0}} (t)} = \ket{\rho_S(t)} \nonumber \\
    &\ket{\rho^{\vec{k} \neq \vec{0}}(0)} = \vec{0}.
\end{align}
The equations are truncated to include $N$ auxiliary density matrices, which together with the number of exponentials $M$ sets the level of approximation. Because the equations only couple adjacent levels of the hierarchy, they can be efficiently integrated to obtain the resulting reduced density matrix, which is the central idea of the HEOM method. In Ref.\citenum{link2024} it was shown that HEOM can be recasted into the PT formalism. This is done by embedding the auxiliary density matrices into an extended Liouville space, 
\begin{align}
    \ket{R} = \sum_n \ket{\rho^{\vec{n}}} \otimes \ket{\vec{n}},
\end{align}
with $\dim[\ket{R}] = S^4 N$. The dynamics of the extended density matrix are given by
\begin{align}\label{eq:extended_heom}
     \partial_t \ket{R} =& \mathcal{L}_S \ket{R} + i \vec{\gamma} \cdot \vec{\hat{N}} \ket{R} \nonumber\\
     &- i \sum_{k=1}^M \left( S \otimes I - I \otimes S \right) A_k^\dagger \ket{R} \nonumber \\
   & - i \sum_{k=1}^M \hat{N}_k \left( \alpha_k S \otimes I - I \otimes \tilde{\alpha}_k S \right) A_k \ket{R} \nonumber \\ 
& = \mathcal{L}_S \ket{R} + \mathcal{L}_{int} \ket{R}, 
\end{align}
where we introduced the number operators $\hat{N}_k$, as well as the raising and lowering operators $ A_k^\dagger,A_k$. Eq.\eqref{eq:extended_heom} can be integrated, and the reduced density matrix can be obtained as
\begin{align}\label{eq:heom-pt}
    &\bra{\mu_N}\ket{\rho_S(N\Delta t)} = \bra{\mu_N,0}\ket{R(N \Delta t)} \nonumber\\
    =&\bra{0}\mathcal{F}^{\{\alpha,\beta\}}\sum_{\mu} \left( \prod_k U_{k}^{\mu_{k+1}\beta_k} U_k^{\alpha_k\mu_{k-1}} \right) \bra{\mu_0}\ket{R(0)} \nonumber \\
    =&\bra{0}\mathcal{F}^{\{\alpha,\beta\}}\ket{0} \sum_{\mu} \left( \prod_k U_{k}^{\mu_{k+1}\beta_k} U_k^{\alpha_k\mu_{k-1}} \right) \bra{\mu_0}\ket{\rho_S(0)} ,
\end{align}
where $\alpha,\beta$ denote indices in the system space and
\begin{align}
    \mathcal{F} = \prod_{n=1}^T e^{\mathcal{L} \Delta t}
\end{align}
is a PT with bond dimension given by the number of auxiliary density matrices $N$. Solving HEOM is therefore equivalent to constructing the process tensor $ \mathcal{F} $, noting that the result needs to be projected back into the physical space to obtain the reduced system density matrix. The tensor diagram corresponding to Eq.~\eqref{eq:heom-pt} is shown in Fig.~\ref{fig:heom-diagram} . 

Because the PT here is obtained without compression, the tensors at each time step are all the same. This has the advantage that only one tensor needs to be computed and saved, at the expense of having a larger bond dimension. This advantage is also available in conjunction with the compression, using a tensor-network contraction scheme which produces a time-translationally invariant PT-MPO\cite{link2024}. This feature is particularly beneficial for simulations with large numbers of time steps. On the other hand, when the PT-MPO is compressed using the J\o rgensen-Pollock scheme~\cite{jorgensen2020} in OQuPy~\cite{oqupaper} one finds that the bond dimension is reduced at the beginning and the end of the process, where memory effects are less important. 

The tensors in the PT produced by HEOM are sparse because of the hierarchical structure, and this can be exploited to reduce the scaling in their contraction from $O(N^2)$ to $O(N)$. However, the number of auxiliary density matrices which are used for typical HEOM calculations~\cite{tang2015,huang2023,krug2023,mangaud2023} is considerably larger than the bond dimensions which arise, in our experience, when the PT-MPO is constructed using SVD compression schemes. This is perhaps not unexpected, because HEOM assumes a rather specific representation of the environment, which or may not be efficient depending on the problem being considered. 

Furthermore, in certain cases the truncation of the HEOM hierarchy leads to instabilities. This is the case when using low temperature, as many Matsubara modes need to be included in the exponential decomposition of the bath correlation function \cite{dunn2019,krug2023}. HEOM has been used to implement optimal control \cite{mangaud2018}, and there have been many extensions to the method, such as better ways to implement low temperature \cite{Fay2022,krug2023}, or using tensor network techniques to improve efficiency \cite{shi2018,mangaud2023,ke2023}.

\subsection{Stochastic methods}

An alternative class of approaches, to the ones discussed so far, are the various stochastic methods for the study of structured baths. Some of these methods are  very similar to each other, if not outright equivalent \cite{wu2018,yan2018,xu2023}. We discuss two of the most commonly applied methods to illustrate the stochastic approach.

\subsubsection{Stochastic Liouville}

Based on early work by Anderson \cite{w.anderson1954} and Kubo \cite{kubo1954}, the stochastic Liouville (SL) method has been used extensively \cite{grabert1984,dattagupta1987,kampen2007}. The method accounts for the effect of the environment by coupling the system to a stochastic field, $\Omega(t)$, as
\begin{align}
    \hat{H}(t) = \hat{H}_S + \hat{S}\,  \Omega (t).
\end{align}
The properties of the environment are then encoded in the statistics of the stochastic field. This can be used as a phenomenological model \cite{grabert1984}. However, assuming that $\Omega(t)$ is a Gaussian-Markovian process characterized by the correlation
\begin{align}
    \expval{\Omega(t)\Omega(\tau)} = \sum_{j=1}^M \alpha_j e^{i \gamma_j (t-\tau)},
\end{align}
the SL method reproduces HEOM \cite{tanimura2006}, meaning that the method can also be used to fully replicate the effect of a particular environment.

\subsubsection{Stochastic Liouville-von Neumann}

\begin{figure}[t]
\centering
\begin{tikzpicture} [
    triangle/.style = {fill=white, isosceles triangle, isosceles triangle apex angle=90, draw=blue!20!black,line width=0.5mm, rounded corners=0.1cm
    },
    border rotated/.style = {shape border rotate=270},
    propagator/.style = {fill=white,circle,draw=white,line width=0.4mm},
    mpo/.style = {fill=red!20,rectangle,draw=red!70!black,line width=0.4mm,rounded corners=0.1cm},
    circlepink/.style = {fill=purple!20!white, circle, isosceles triangle apex angle=90, draw=blue!20!black,line width=0.5mm, rounded corners=0.1cm
    },
    half circle/.style={fill=white,semicircle,shape border rotate=180,anchor=chord center,  draw=blue!20!black,line width=0.5mm, rounded corners=0.02cm
      }
  ]


    \def \xa {0.0}
    \def \xb {0.8}
    \def \deltay {1.0}

    \node (alabel) at (-0.8,4.5) {$(a)$} ;
    \node (mu) at (\xb,\deltay*3.5+0.5*\deltay) {$\mu$} ;
    \node (nu) at (\xb,\deltay*3.5-0.5*\deltay) {$\nu$} ;

    \node[mpo,minimum size=0.4cm,label = left:{}]   (mpoa) at (\xa,\deltay*3.5) {\small };
    \draw [line width = 0.4mm] 
    (mpoa) -- (mu)
    (mpoa) -- (nu);

    \node (eq1) at (3*\xb,\deltay*3.5) {$=U_{SB}^{\mu\nu}(\xi_k^i,\nu_k^i) $};


    \node (arrow) at (2*\xb,\deltay*2.5) {$\downarrow$};

    \def \xa {0.0}
    \def \xb {0.8}
    \def \deltay {1.0}

    \node (alabel) at (-0.8,4.5) {$(a)$} ;
    \node (mu) at (\xb,\deltay*1.5+0.5*\deltay) {$\mu$} ;
    \node (nu) at (\xb,\deltay*1.5-0.5*\deltay) {$\nu$} ;
    \node (i) at (\xa,\deltay*1.5+0.7*\deltay) {$i$} ;
    \node (j) at (\xa,\deltay*1.5-0.7*\deltay) {$j$} ;

    \node[mpo,minimum size=0.4cm,label = left:{}]   (mpoa) at (\xa,\deltay*1.5) {\small };
    \draw [line width = 0.4mm] 
    (mpoa) -- (mu)
    (mpoa) -- (nu);

    \draw [line width = 0.6mm, draw=red!70!black] 
    (mpoa) -- (i)
    (mpoa) -- (j);

    \node (eq1) at (3.4*\xb,\deltay*1.5) {$=U_{SB}^{\mu\nu}(\xi^i_k,\nu^i_k) \delta_{ij}$};

    \node (arrow) at (2*\xb,\deltay*2.5) {$\downarrow$};


    \def \ptx {5.5}
    \def \pty {0.0}
    
    \def \xa {0.0}
    \def \xb {0.8}
    \def \deltay {1.0}

    \node (blabel) at (\ptx-1.0,\pty+4.5) {$(b)$} ;
    \node[triangle, border rotated, minimum size=0.02cm, fill=white, draw=white] (initial_state) at (\ptx+\xb,\pty-0.0) {};

    \node[half circle, minimum size=0.02cm] (initial_cap) at (\ptx+\xa,0.5) {};
    \node[half circle, rotate=180,minimum size=0.02cm] (final_cap) at (\ptx+\xa,\deltay-0.5+\deltay*3+\deltay) {};

    \foreach \n in {0,1,2,3}{

    \node[propagator, minimum size=0.1cm] (prop\n) at (\ptx+\xb,\deltay*\n+0.5*\deltay) {\tiny \tiny };
    \node[mpo,minimum size=0.4cm] (mpo\n) at (\ptx+\xa,\deltay*\n+\deltay) {\small };
    \node[propagator, minimum size=0.1cm] (prop5) at (\ptx+\xb,\deltay*4+0.5*\deltay) {\tiny \tiny };
    }

    \draw [line width = 0.4mm] 
    (initial_cap) -- (mpo0)
    (prop0) -- (mpo0)
    (mpo1) -- (mpo2) ;

    \foreach \n [evaluate=\n as \m using int(\n-1)] in {1,2,3}{
    \draw [line width = 0.6mm, draw=red!70!black] 
    (mpo\m) -- (mpo\n);
    \draw [line width = 0.4mm]
    (mpo\m) -- (prop\n)
    (prop\n) -- (mpo\n);
    ;}

    \draw [line width = 0.4mm] 
    (mpo3) -- (prop5)
    (mpo3) -- (final_cap)
    ;


\end{tikzpicture}
\caption{(a) Time-evolution operator for the Markovian bath, which is transformed into a 4 legged-tensor, by indexing the noise configurations and adding a Kronecker delta. (b) The physical process tensor is obtained by contracting the tensors and tracing over the external legs, represented by the caps. The bond dimension of the PT corresponds to the number of noise configurations considered.} \label{fig:stochastic-diagram}
\end{figure}
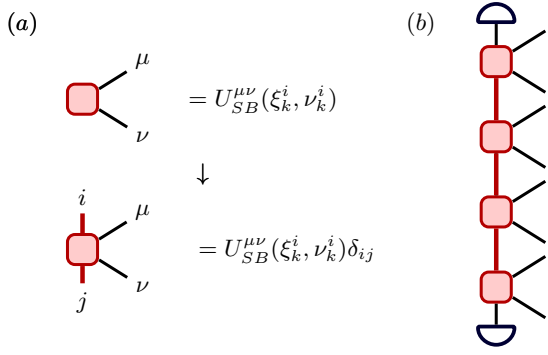

The stochastic Liouville-von Neumann \cite{stockburger1999,stockburger2001,stockburger2002, stockburger2004} (SLVN) method takes a different starting point, based on the influence functional $\mathcal{F}(s,s')$ for a system coupled to a bath of harmonic oscillators. Here $s,s^\prime$ represent coordinates of the system associated with the forward (backward) part of the time evolution of the density matrix. Performing a change of variables, $x= s-s',r=(s+s')/2$, the influence functional can be rewritten as a Gaussian integral over two complex auxiliary fields,
\begin{align}
    \mathcal{F}(x,r) =& \int \mathcal{D}[\xi]\int \mathcal{D}[\nu] W[\xi,\xi^*,\nu,\nu^*] \nonumber \\
    &\cross \exp[i\int_{t_0}^{t}d\tau (\xi(\tau) x(\tau) + i \nu(\tau)r(\tau))].
\end{align}
The probability distribution $W[\xi,\xi^*,\nu,\nu^*]$ is fixed by imposing
\begin{align}
    &\expval{\xi(t)\xi(t')]}_W = \Re[C(t-t')] \nonumber \\
    &\expval{\xi(t)\nu(t')}_W = 2i\theta(t-t')\Im[C(t-t')]  \nonumber\\
    &\expval{\nu(t)\nu(t')}_W = 0,
\end{align}
where the averages are taking with respect to the probability distribution $W$, and $C(t-t')$ is the bath correlation function. We can now obtain the reduced density matrix as the average of stochastic density matrices,
\begin{align}
    \rho_S(t) = \expval{ \rho(\xi,\nu,t) }_W,
\end{align}
where the stochastic system density matrices evolves according to the Liouvillian
\begin{align}
    i\dot{\rho}(\xi,\nu,t) =& [H_0,\rho(\xi,\nu,t)] \\
    & - \xi(t)[S,\rho(\xi,\nu,t)] - \frac{1}{2}\nu(t)\{S,\rho(\xi,\nu,t) \}. \nonumber
\end{align} Here $S$ is the system operator which appears in the system-bath coupling, and is $\sigma_z/2$ in the case of Eq.~\eqref{eq:sb-model}. Other stochastic methods, such as non-Markovian quantum state diffusion \cite{strunz1996,diosi1997,diosi1998,strunz1999} and stochastic decoupling \cite{yan2018,wu2018} follow similar approaches.

The resulting Liouvillian can be decomposed into a system term $\mathcal{L}_S$ and an interaction term $\mathcal{L}_{SB}$, describing the Markovian evolution of the stochastic density matrix. All memory effects are encoded in the correlation of the stochastic fields and they become evident when we perform the average over noise configurations. Suppose that the noise configurations are indexed by a variable $i$. For a particular configuration, Trotterization results in the step
\begin{align}
    \rho^{\mu_{k}}_{k}(\xi^i,\nu^i) = \sum_{\mu_{k-1}\nu_k }U_{SB}^{\mu_{k}\nu_k}(\xi^i_k,\nu^i_k) U_S^{\nu_k \mu_{k-1}} \rho_{k-1}^{\mu_{k-1}},
\end{align} where the arguments of $U_{SB}$ are the values of the stochastic fields at the time step $k$.
Consider the tensor
\begin{align}
    O[k]^{\mu\nu}_{ij} = U_{SB}^{\mu \nu} (\xi^i_k,\nu^i_k) \delta_{ij},
\end{align}
shown in Fig.~\ref{fig:stochastic-diagram}(a), where the additional legs label the particular noise configuration. The physical PT is obtained by averaging over noise configurations. This can be done by contracting the bond legs of the tensors $O[k]$, and summing over the first and last leg, as shown diagramatically in Fig.~\ref{fig:stochastic-diagram}(b). In this way we can construct a PT-MPO representation of the stochastic Liouville equation, with a constant bond dimension which corresponds to the number of trajectories sampled.

The bond dimension of the resulting physical PT depends on the number of noise configurations, or trajectories, averaged. The efficiency of the method therefore heavily depends on how the trajectories are sampled. This is usually done randomly, as there is no way of knowing a priori which trajectories are the most relevant ones. From this perspective, the observation that the SVD compression algorithms lead, in many cases, to PT-MPOs with small bond dimensions, may be interpreted in terms of them selecting the most relevant trajectories.

Furthermore, stochastic simulations typically suffer from instabilities, particularly in the long time limit. This is why other dynamical methods, such as HEOM or PT-based methods, have been favoured recently \cite{wu2018}. Nonetheless, stochastic methods can prove useful in the short-time regime, or complement other methods in a hybrid approach \cite{moix2013,yan2022}, and have also been used in quantum optimal control \cite{schmidt2011}.

\subsection{Augmented systems}

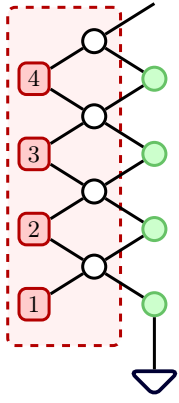
\begin{figure}[t]
\centering
\begin{tikzpicture} [
    triangle/.style = {fill=white, isosceles triangle, isosceles triangle apex angle=90, draw=blue!20!black,line width=0.5mm, rounded corners=0.1cm
    },
    border rotated/.style = {shape border rotate=270},
    propagator/.style = {fill=green!20,circle,draw=green!60!black!60,line width=0.4mm},
    propagatorA/.style = {fill=white,circle,draw=black,line width=0.4mm},
    mpo/.style = {fill=red!20,rectangle,draw=red!70!black,line width=0.4mm,rounded corners=0.1cm},
    circlepink/.style = {fill=purple!20!white, circle, isosceles triangle apex angle=90, draw=blue!20!black,line width=0.5mm, rounded corners=0.1cm
    },
  ]
    \def \xa {0.0}
    \def \xb {0.8}
    \def \deltay {1.0}
    
    \node[triangle, border rotated, minimum size=0.02cm] (initial_state) at (2*\xb,-0.0) {};

    \node[fill=red!5,dashed, rectangle,draw=red!70!black,line width=0.4mm,rounded corners=0.1cm,minimum height=4.5cm,minimum width=1.5cm ] (ptbox) at (0.5*\xb,\deltay*1.5+\deltay+0.2) {};
    
    \foreach \n  [evaluate=\n as \m using {int(\n+1)}] in {0,1,2,3}{

    \node[propagator, minimum size=0.1cm] (prop\n) at (2*\xb,\deltay*\n+\deltay) {\tiny \tiny };
    \node[mpo,minimum size=0.4cm] (mpo\n) at (\xa,\deltay*\n+\deltay) {\small \m};
    
    }

    \foreach \n in {0,1,2,3}{

    \node[propagatorA, minimum size=0.1cm] (propA\n) at (\xb,\deltay*\n+1.5*\deltay) {\tiny \tiny};

    }
    \draw [line width = 0.4mm] 
    (initial_state) -- (prop0)
    (prop0) -- (propA0)
    (propA0) -- (mpo0);
      
    \foreach \n [evaluate=\n as \m using int(\n-1)] in {1,2,3}{
    \draw [line width = 0.4mm]
    (prop\n) -- (propA\n)
    (prop\n) -- (propA\m)
    (mpo\n) -- (propA\m)
    (propA\n) -- (mpo\n);
    ;}

    \draw [line width = 0.4mm] 
    (propA3) -- (2*\xb,\deltay*4+\deltay)
    ;

\end{tikzpicture}
\caption{Diagram for the system density matrix evolution. Open (white) circles correspond to propagators of the auxiliary system. Bath MPOs do not couple to each other due to Markovianity, but do so through the auxiliary system. The shaded region inside the dashed line corresponds to the process tensor of the combined bath plus auxiliary degrees of freedom. } \label{fig:augmented-diagram}
\end{figure}

The effect of structured environments can also be modeled by coupling the system to auxiliary modes which are in turn coupled to a Markovian bath \cite{garraway1997,tamascelli2018,basilewitsch2019,pleasance2020}. The enlarged system composed by the physical system and the auxiliary modes evolves as
\begin{align}
    \rho_{S+A} = (\mathcal{L}_S (\vec{u}) + \mathcal{L}_{SA} + \mathcal{L}_A )\rho_{S+A},
\end{align}
where $\mathcal{L}_S(\vec{u})$, containing the control parameters, acts solely on the system, $\mathcal{L}_{SA}$ includes the interaction of system and auxiliary modes, and $\mathcal{L}_{A}$, acting solely on the auxiliary modes, includes a Lindblad dissipator modelling the effect of a Markovian bath. After Trotterization, we have
\begin{align}
    \rho_{S+A,T} = \left(\prod_{k=1}^{T}  U_{A}U_{SA}U_{S}(t_k) \right)  \rho_{S+A,0}.
\end{align}
By defining $O = U_A U_{SA}$, we have
\begin{align}
    \rho^{\mu_{T} \alpha_{T}}_{S+A,T} = \sum_{\substack{\alpha_0 ... \alpha_{T-1}\\ \{\mu\},\{\nu\}}}  \left( \prod_{k=1}^{T}  O^{\mu_{k}\nu_{k}}_{\alpha_{k}\alpha_{k-1}}  U^{\nu_k \mu_{k-1}}_{S}(t_k) \right)  \rho^{\mu_0 \alpha_0}_{S+A,0},
\end{align}
where $\mu,\alpha$ denote the system and auxiliary degrees of freedom, respectively. After taking the trace over the auxiliary degrees of freedom, we have
\begin{align}
    \rho^{\mu_{T}}_{S,T} =&  \sum_{\substack{\{\alpha\},\\ \{\mu\},\{\nu\}}}  \left( \prod_{k=1}^{T}  O^{\mu_{k}\nu_{k}}_{\alpha_{k}\alpha_{k-1}}  U^{\nu_k \mu_{k-1}}_{S}(t_k) \right)  \rho_{S+A,0}^{\mu_0 \alpha_0} \nonumber \\
    =&\sum_{\{\mu\},\{\nu\}}\left( \sum_{\{\alpha\}} \prod_{k=1}^{T}  O^{\mu_{k}\nu_{k}}_{\alpha_{k}\alpha_{k-1}} \rho_{A,0}^{\alpha_0} \right) \nonumber \\ &\qquad \qquad\times \left( \prod_{k=1}^{T}  U^{\nu_k \mu_{k-1}}_{S}(t_k) \right)  \rho^{\mu_0}_{S,0} \nonumber \\
    =&\sum_{\{\mu\},\{\nu\}}\mathcal{F}^{\vec{\mu}\vec{\nu}}  \left( \prod_{k=1}^{T}  U^{\nu_k \mu_{k-1}}_{S}(t_k) \right)  \rho^{\mu_0}_{S,0} .
\end{align}
That is precisely the expression for a process tensor with bond dimension equal to the dimensionality of the auxiliary space. The corresponding diagram is shown in Fig.~\ref{fig:augmented-diagram}.

The usefulness of the method depends on how efficient the representation of the bath is in terms of the auxiliary modes. This decomposition is not unique, and the methods do not ensure that a minimal representation is obtained. Nonetheless, once the process tensor is constructed in terms of the auxiliary modes, tensor network techniques can be employed to further compress the process tensor.

\section{Generalizations in the process tensor framework}

In the above we have focused on the problem of determining a time-dependent Hamiltonian, or more generally sequence of unitary operations, that steers a system towards a target state at a particular time. A powerful feature of the process tensor, however, is that it describes a general process in which a quantum system is subject to interventions. This structure makes it straightforward to consider more general problems, such as the optimization of trajectories, and the inclusion of measurements and feedback in the control \cite{pollock2018}.  

Since we focused on state transfer, we only looked at terminal costs. Although less common in quantum optimal control, in some applications it might be beneficial to also include running costs. For example, if we want to minimize the strength of the control operations. Running costs are less straightforward to deal with than terminal costs within the PT framework, however, it would be possible to adapt the approach used in classical optimal control of introducing an additional variable which tracks the running cost \cite{cesari1983}. The complication is that the propagation of this additional variable may propagate nonlinearly.

One can also be interested in optimizing quantities related to the environment, such as the heat currents flowing between system and bath. For Gaussian environments  (like the ones implemented in OQuPy), any multi-time correlator of the environment can be obtained from the correlations in the system \cite{gribben2022}, which can be obtained efficiently using the PT. For other general types of environment, counting fields can be introduced in the calculation of the process tensor to obtain the desired quantities \cite{popovic2021}. 

Finally, new methods are being developed within the framework of the process tensor. One example is the periodic-boundary PT obtained adapting the iTEBD algorithm used in MPS \cite{link2024}. Combining the optimization methods described here with this algorithm would allow one to more efficiently optimize the controls needed to reach a target steady-state, without the need to simulate the dynamics for large number of time-steps.

\section{Conclusions}

In this paper we analyzed some of the most common methods used to simulate open quantum systems coupled to structured environments. By expressing the different methods in terms of process tensors it is possible to compare the efficiency of the methods, and in particular their suitability for optimization calculations, directly by comparing their bond dimension. In the case of TDVP, one cannot construct an equivalent process tensor, and therefore we analyzed the implementation of the adjoint method independently, in order to compare with the rest of the methods. 

All methods rely on reducing the effective dimensionality of the environment. The methods which use SVD compression schemes to compute the PT-MPOs have the advantage that they can, and in our experience often do, numerically discover an efficient representation of the environment. The other approaches make more assumptions about how to represent the environment, suggesting they will not perform as well in general, although they may perhaps perform better in specific cases. Furthermore, two of these other methods present particular advantages that are worth considering. At the expense of accuracy, we find the the simplicity of the TDVP approach can reduce considerably the cost of doing optimization. This can be useful when qualitative results are sufficient. HEOM also provides several advantages. While generally resulting in a larger bond dimension compared to PT-MPO, the hierarchical structure could be exploited to reduce the scaling cost of the contraction, and could result in similar efficiency to PT-MPO if tensor network compression techniques are implemented on top of the conventional approach.

In order to meet institutional and research funder open access requirements, any author accepted manuscript arising shall be open access under a Creative Commons Attribution (CC BY) reuse licence
with zero embargo.

\begin{acknowledgments}
We thank Brendon Lovett for helpful discussions and feedback on the manuscript, and acknowledge support from Science Foundation Ireland (21/FF-P/10142).

\end{acknowledgments}

\bibliography{bibliography}

\end{document}